  \documentclass[onecolumn,usenatbib]{mn2e}
\usepackage{epsf}
\usepackage{epsfig}
\newcommand{\simgt}{\lower.5ex\hbox{$\; \buildrel > \over \sim \;$}}
\newcommand{\simlt}{\lower.5ex\hbox{$\; \buildrel < \over \sim \;$}}

\begin{document}
%%%%%%%%%%%%%%%%%%%%
%% Title page
%%%%%%%%%%%%%%%%%%%%
\title[Evolution of the $M_{BH}$-$M_{sph}$ relation]
{Mass-dependent evolution of the relation
between supermassive black hole mass and host 
spheroid mass since z $\sim$ 1 }

\author[S. Kisaka and Y. Kojima]
{Shota Kisaka\thanks{E-mail: kisaka@theo.phys.sci.hiroshima-u.ac.jp}
and
Yasufumi Kojima\thanks{E-mail: kojima@theo.phys.sci.hiroshima-u.ac.jp}
\\
Department of Physics, Hiroshima University, Higashi-Hiroshima, Hiroshima
739-8526, Japan}
\maketitle
%%%%%%%%%%%%%%%%%%%%%%%%%%%%%%%%%%%%%%%%%%%%%%%%%%%%%%%%%%%%%%%%%%%%%%
\begin{abstract}
  We investigate the evolution of supermassive black hole mass 
($M_{BH}$) and the host spheroid mass ($M_{sph}$)
in order to track the history of the $M_{BH}$-$M_{sph}$ relationship.
The typical mass increase of $M_{BH}$ is calculated by  
a continuity equation and accretion history, which
is estimated from the active galactic nucleus (AGN) luminosity function. 
The increase in $M_{sph}$ is also calculated by using
a continuity equation and a star formation model, 
which uses observational data for
the formation rate and stellar mass function. 
We find that the black hole to spheroid mass ratio is expected to 
be substantially unchanged since $z\sim1.2$ for high mass objects 
($M_{BH}>10^{8.5}M_{\odot}$ and $M_{sph}>10^{11.3}M_{\odot}$). 
In the same redshift range, the spheroid mass is found to increase more 
rapidly than the black hole mass if $M_{sph}>10^{11}M_{\odot}$.
The proposed mass-dependent model is consistent with
the current available observational data in the $M_{BH}$-$M_{sph}$
diagram.
\end{abstract}
%
%%%%%%%%%%%%%%%%%%%%%%%%%%%%%%%%%%%%%%%%%%%%%%%%%%%%%%%%%%%%%%%%%%%%%%

\section{INTRODUCTION}

  Recent observations show that most, if not all, nearby massive spheroids
(elliptical galaxies, lenticular and spiral bulges) have 
supermassive black holes(SMBHs) with masses in the range 
of $M_{BH}=$ $10^{6}$-$10^{10}M_{\odot}$ at their center. 
In the local universe, the SMBH mass is correlated with characteristic
parameters of the host spheroid: the luminosity $L_{sph}$\citep{KR95,
MH03, G07}, the mass $M_{sph}$ of the spheroid \citep{Ma98, HR04},
the stellar velocity dispersion $\sigma$\citep{Ge00, FM00, Tr02} and
the various other properties (e.g., \citet{Gr07, AR07, KKO08}).
Understanding the origin of these relations is thought to provide 
insights into the "co-evolution" of SMBHs and the host spheroids. 
The role of central black holes in galaxy formation 
and evolution is not yet established 
since the SMBH mass is tiny compared with the galaxy as a whole. 
Areas that are understood at present, as well as those suggested 
for future observation, were recently reviewed in \citet{Ca09}.
Some tentative
theories have been proposed to clarify the origin of these
correlations (e.g., \citet{SR98}), and they are fairly successful at
reproducing the correlations at the present time.
However, no theoretical model is widely accepted since
there are many unknown parameters in the evolutionary model.
For example, in semianalytic models,
depending on the model assumptions,
the evolution of the scaling relation differs markedly
(e.g., \citet{C06, Ho09}). 
Therefore, it would be of value to study the evolution 
using an alternative approach.

Cosmic evolution of the ratio $M_{BH}$/$M_{sph}$
is observationally studied.
\citet{Mc06} used the 3C RR sample of radio-loud active galactic
nuclei (AGNs) for $0<z<2$.
The number of sources for each redshift bin is not sufficiently large to
examine the $z$-dependence.
By fitting data for the relation $M_{BH}$/$M_{sph}\propto(1+z)^{\gamma}$,
the index is estimated as $\gamma = 2 $, although 
data for $ z<1$ are also consistent with no evolution ($\gamma = 0 $).
\citet{Tr07} also derived a similar result
for SMBH masses in a sample of 20 Seyfert galaxies
around $z=0.36$; however, the evolution was slightly weak,  
$M_{BH}$/$M_{sph}\propto(1+z)^{1.5}$.
It is not easy at present to judge whether or not cosmological evolution
exists, due to several selection biases \citep{La07}.

  The evolution of the correlations can also be inferred from the
demographics of galaxies and AGNs \citep{MRD04}. 
In some reasonable theories, the nuclear activity of AGNs
is proportional to the mass accretion rate into the SMBHs.
The luminosity function of AGNs and the redshift evolution
measure the buildup of SMBH mass (e.g., \citet{S82}).
The evolution of the galaxies is related to the star formation 
history.
\citet{MRD04} combined
the evolution of the SMBH mass density derived from the mass
accretion history with that of the stellar mass density derived from
the star formation history of the Universe, 
in order to investigate the evolution of the 
$M_{BH}$-$M_{sph}$ relation. 
They found that the growth of SMBHs appears to
predate that of spheroids as $M_{BH}$/$M_{sph}\propto(1+z)^{0.4-0.8}$,
for which the power index of the redshift factor is
smaller than that obtained by observations.

Here, we consider the evolutionary model by
taking into account the mass-dependent growth in the phenomenological
analysis. That is, a simple power-law relation 
$M_{BH}$/$M_{sph}\propto(1+z)^{\gamma}$, is not assumed.
Only integrated quantities have been used so far in phenomenological 
studies (e.g., \citet{MRD04}). 
Our model is an improvement of previous models, in which the 
mass-dependent effect is now included as an important factor.
The mass-dependent property of cosmological evolution is 
known as "downsizing", 
in which star formation becomes active in
less massive galaxies as the cosmic time increases
(e.g., \citet{Co96}). 
Similarly, an increase in the number of
AGNs with redshift compared with the local universe is followed by a
decline beyond a peak whose redshift depends on luminosity 
and hence {\bf on} the SMBH mass, assuming a constant Eddington 
ratio (e.g., \citet{Ye09}).
More recently, \citet{La09} have studied the mass dependence of redshift 
for active black holes by using the Malmquist-bias-unaffected method, 
and argued that AGN samples at lower redshifts are increasingly dominated by
less massive black holes.
These observations suggest that
the evolution of both SMBH mass and galaxy mass depends on their masses.
In this paper, we extend the evolutionary models
based on the AGN luminosity function and star formation 
history by including the mass-dependent effect
and derive the evolution of the $M_{BH}$-$M_{sph}$ relation
at certain cosmic ages.

  We organize the paper as follows. In Section 2, 
we briefly review the evolutionary model of SMBHs
considered by \citet{Ma04, SWM09}, in order to
clarify the model parameters.
We assume a mean Eddington ratio for
the active SMBHs. By adopting a continuity equation 
and an assumed initial condition, 
the AGN luminosity function correlates directly with the SMBH mass 
function at all times. In Section 3, we describe the evolution
model of spheroids. Using a method similar to that used for SMBHs, that is, 
a continuity equation and an
assumed initial condition, the stellar mass function and specific star
formation rate are used to calculate the mass increase of the
spheroids. 
In Section 4, we combine these two evolutionary models
and discuss the evolution of the $M_{BH}$-$M_{sph}$ relation. Finally
in Section 5, we provide a summary. Throughout the paper,
we adopt a cosmology with $\Omega _{m}=0.3$, $\Omega _{\Lambda }=0.7$
and $H_{0}=70$ km s$^{-1}$ Mpc$^{-1}$.

\section{ EVOLUTIONARY MODEL OF BLACK HOLES}

  In the following section, we review the evolutionary equation of the SMBH mass function
according to \citet{Ma04, SWM09}. 
The mass of black holes generally
increases through mass accretion, which also causes AGN
activity. The SMBH mass function is therefore determined from the observed
luminosity function of AGNs (e.g., \citet {CMW71, SB92, Ma04,SWM09}). 
We assume that AGNs are powered by accretion into SMBHs, and
that SMBH growth takes place during a phase in which the
AGN is shining at a fraction $\lambda $ of the Eddington luminosity, 
converting the mass accretion rate with a radiative efficiency 
$\epsilon $.
The number density of SMBHs with mass $M_{BH}$ in co-moving space is
denoted by $n(M_{BH},t)$, which satisfies the continuity equation 
\begin{equation}
\frac{\partial n(M_{BH},t)}{\partial t}+\frac{\partial }{\partial
M_{BH}} [n(M_{BH},t)\langle \dot{M}_{BH}(M_{BH},t)\rangle ]=0,
\label{2.4} 
\end{equation}
where $\langle \dot{M}_{BH}(M_{BH},t)\rangle $ represents the mean 
mass-growth rate of $M_{BH}$ at time $t$.
The mass growth is related to AGN activity and hence
is expressed by the bolometric luminosity function of the AGNs.
\citet{Ma04} derived an evolutionary equation of
the SMBH number as
%%%%%%%%%%%%%%%%%%%%%%
\begin{equation}
\frac{\partial n(M_{BH},t)}{\partial t}=-\frac{(1-\epsilon )\lambda
^{2}c^{2}}{\epsilon t_{Edd}^{2}\ln 10}\biggl[\frac{\partial \Phi
(L_{bol},t)}{\partial L_{bol}}\biggr],  \label{2.6}
\end{equation}
where $t_{Edd}$ is the Eddington time.

  The mass growth of SMBHs between redshifts
$z_{1}$ and $z_{2}$ is calculated by the integration 
\begin{eqnarray}
M(z_{2})-M(z_{1}) &=&\int \langle \dot{M}_{BH}(M_{BH},t)\rangle
\frac{dt}{dz}dz  \nonumber \\
&=&\frac{(1-\epsilon )\lambda }{\epsilon t_{Edd}\ln 10}\int 
\frac{1}{n(M_{BH},t)}[\Phi (L_{bol},t)]\frac{dt}{dz}dz.  \label{2.7}
\end{eqnarray}

  The evolutions of SMBH mass density and typical mass at $z$ are 
calculated by using 
eqs. (\ref{2.6}) and (\ref{2.7}), which are expressed by the AGN
luminosity function with two parameters $\epsilon $ and $\lambda $.
We adopt 
the AGN luminosity function of \citet{SWM09} 
with a smoothing modification by \citet{RF09}.
The observed AGN luminosity function is  
converted to the bolometric luminosity by using the 
luminosity-dependent correction of \citet{Ma04}. 
The initial condition for the integration of eq. (3)
is chosen as the condition at $z=6$, which is  
the same as \citet{SWM09}. 
They noted that the integration does
not significantly depend on the initial value after $z\sim 3.5$
unless the SMBH duty cycle is extremely small.

  In order to constrain the free parameters $\epsilon $ and $\lambda $, 
we compare the local SMBH mass function derived from the evolution model with 
an observational value. \citet{SWM09} carried out a detailed comparison 
by using the SMBH mass function derived by various methods.
We also obtained the same results as \citet{SWM09},
even when including the technical smoothing modification.
The best-fitting values
are $\epsilon =0.065$ and $\lambda =0.42$.
Uncertainty in the radiative efficiency $\epsilon $
is not a highly important parameter affecting the mass growth
of SMBHs since it affects only the overall normalization.
The parameter is fixed as $\epsilon =0.065$ from now on.
The Eddington ratio $\lambda $ is a critical parameter
describing SMBH evolution. 
\citet{SWM09} estimated a reasonable range of their model parameter,
which is converted as $\lambda =0.28$-$0.56$.
The range implies not uncertainty in any statistical sense,
but rather a preferable range.

\begin{figure}
 \begin{center}
  \includegraphics[width=110mm, angle=270]{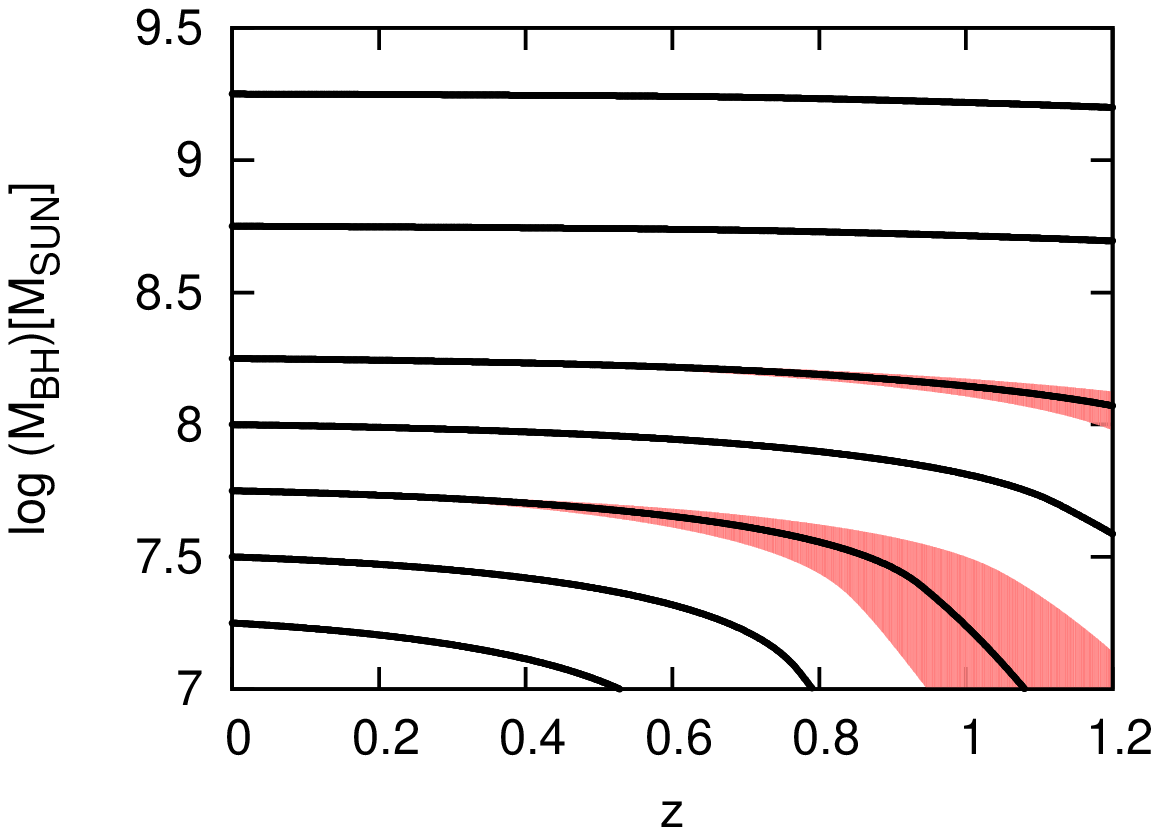}
 \end{center}
 \caption{Averaged growth history of SMBHs computed using AGN 
  luminosity function and $\epsilon=0.065$.
  Eddington ratio is $\lambda=$0.42.
  Red shaded areas show the difference in $\log M_{BH}$ between 
  $\lambda=0.56$ and $\lambda=0.28$  
  for representative tracks at $z=$0.3, 0.5, 0.7, 0.9 and 1.1.}
 \label{fig:2.1}
\end{figure}

In Figure \ref{fig:2.1}, we plot the averaged growth history of SMBHs with
different masses for $0<z< 1.2$.  
The mass range of black holes for which current observational data 
are available is limited to 
$10^{7}<M_{BH}/M_{\odot}<10^{9.5}$. 
 It is clear that the growth depends on
black hole mass.  
A black hole with $M_{BH}\sim 10^{7}M_{\odot }$
rapidly increases its mass from $z=1.2$ to $z=0$, while one with 
$M_{BH}\sim 10^{9}M_{\odot }$ stops its growth at $z=1.2$. 
We also demonstrate the upper and lower limits of
the Eddington ratio 
for two representative evolutionary tracks as shaded areas. 
One is the track of $M_{BH} = 10^{8.5}M_{\odot }$ at
$z=0$, and the other is $M_{BH} = 10^{7.5}M_{\odot }$.
We find that the uncertainty is important around the
bend of the curve, where the growth is rapid.
The uncertainty does not seriously affect the evolution
for $M_{BH} > 10^{8}M_{\odot }$.
The uncertainty in the track of less massive black holes
can be clearly seen, for example,
$\sim 0.4$dex in $\log M_{BH} $ for
$M_{BH} = 10^{7.5}M_{\odot }$ at $z=0.9$.

\section{ EVOLUTIONARY MODEL OF SPHEROIDS}

  In this section, we introduce a simple model of spheroid mass
evolution. 
It is well known that galaxies form two sequences in color-magnitude
space: star-forming, late-type galaxies occupy the blue cloud, whereas
quiescent, bulge-dominated early-type galaxies reside on the red sequence
(e.g., \citet{Be04}). 
We focus on early type galaxies and derive the spheroid mass 
evolution assuming a fixed bulge-to-total ratio (B/T).

  We use the method reported by \citet{Be07} 
to determine the evolution of the stellar mass function. 
The number densities of both the early- and
late-type galaxies, respectively, satisfy the continuity equation with a 
source function $S$ for each type as 
\begin{equation}
\frac{\partial n_{early}(M,t)}{\partial t}+\frac{\partial }{\partial M}
[n_{early}(M,t)\dot{M}_{early}(M,t)]=S_{early}(M,t),  \label{3.1}
\end{equation}
\begin{equation}
\frac{\partial n_{late}(M,t)}{\partial t}+\frac{\partial }{\partial M}
[n_{late}(M,t)\dot{M}_{late}(M,t)]=S_{late}(M,t).  \label{3.2}
\end{equation}
For all galaxies, we have
\begin{equation}
\frac{\partial n_{all}(M,t)}{\partial t}+\frac{\partial }{\partial M}
[n_{all}(M,t)\dot{M}_{all}(M,t)]=S_{all}(M,t),  \label{3.3}
\end{equation}
where 
\begin{equation}
n_{all}(M,t)=n_{early}(M,t)+n_{late}(M,t),  \label{3.4}
\end{equation}
\begin{equation}
n_{all}(M,t)\dot{M}_{all}(M,t)=n_{early}(M,t)\dot{M}
_{early}(M,t)+n_{late}(M,t)\dot{M}_{late}(M,t),  \label{3.5}
\end{equation}
\begin{equation}
S_{all}(M,t)=S_{early}(M,t)+S_{late}(M,t). 
\end{equation}
 \citet{Be07} calculated the evolution of the mass function 
for all galaxies with an
assumption $S_{all}=0,$ and found that the results were consistent with the
observational COMBO-17 data at $0.2<z<1.0$ \citep{Bo06}. 
They also found $\partial n_{late}/\partial t=0$ 
by comparing the observed mass function
of all galaxies with that of late-type galaxies between $z=0$ and $z\sim 0.9$. 
We adopt both conditions in our analysis.
The continuity can be written as
\begin{eqnarray}
\frac{\partial n_{early}(M,t)}{\partial t} &=&-\frac{\partial }{\partial M}
[n_{all}(M,t)\dot{M}_{all}(M,t)]  \nonumber \\
&=&-\frac{\partial }{\partial M}[n_{early}\langle \dot{M}_{early}(M,t)
\rangle ],
\end{eqnarray}
where 
\begin{equation}
\langle \dot{M}_{early}(M,t)\rangle =\frac{n_{all}(M,t)}{n_{early}(M,t)}
\dot{M}_{all}(M,t).  \label{3.7}
\end{equation}
The typical mass of the early-type galaxies is obtained by 
integrating the equation 
\begin{equation}
M_{early}(z_{2})-M_{early}(z_{1})=\int 
\langle  \dot{M}_{early}(M,t)\rangle
\frac{dt}{dz}dz.  \label{3.8}
\end{equation}
The spheroid mass evolution is determined from the stellar
mass function of early-type galaxies adopting B/T=0.7, 
which does not depend on the redshift or galaxy mass 
(see \citet{TOU06}).

\begin{figure}
 \begin{minipage}{0.47\hsize}
  \begin{center}
   \includegraphics[width=47mm, angle=270]{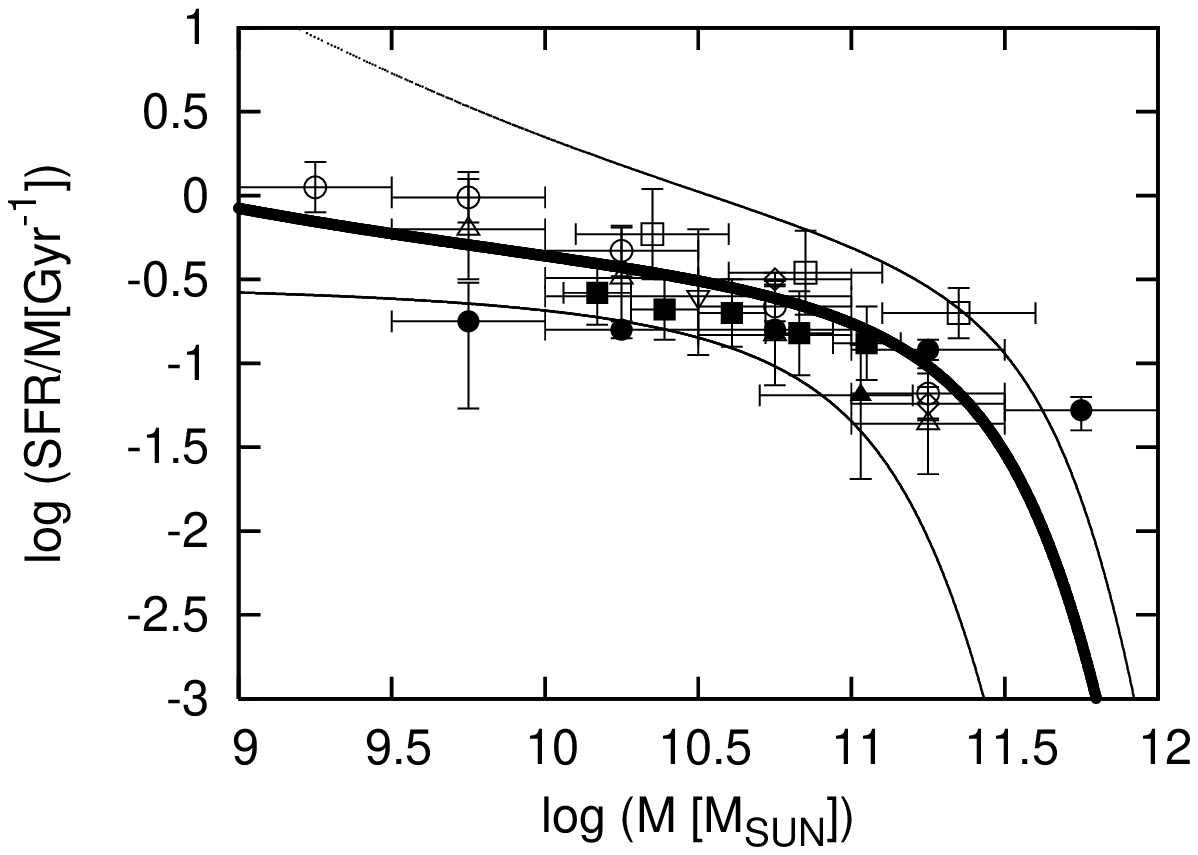}
  \end{center}
 \end{minipage}
 \begin{minipage}{0.47\hsize}
  \begin{center}
   \includegraphics[width=47mm, angle=270]{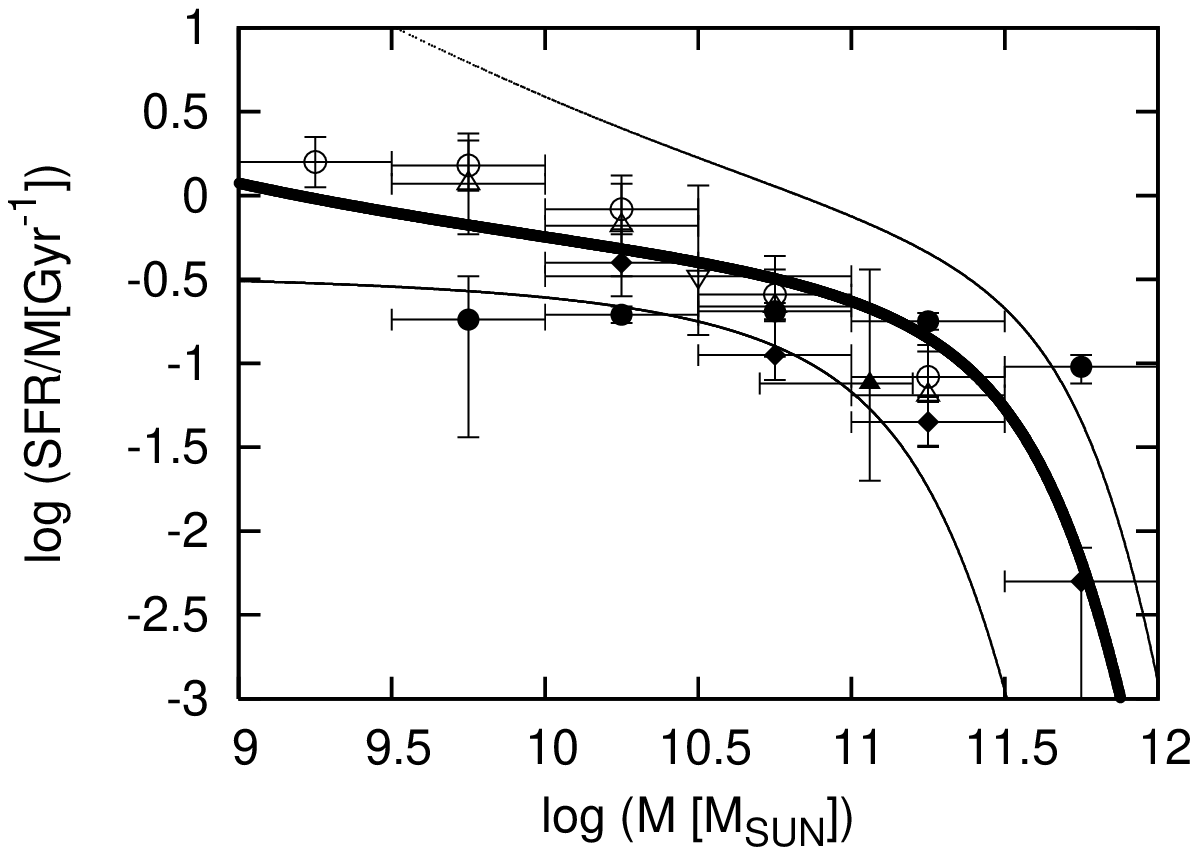}
  \end{center}
 \end{minipage}
 \caption{Specific star formation rate as a function of 
 stellar mass from $z=0.3$ (upper left), 
 $z=0.5$ (upper right), $z=0.7$ (lower left)
 and $z=0.9$ (lower right).
 Thick lines show staged $\tau$ models 
 and thin-lines show
 effect of varying $c_{\alpha}$ and $c_{\beta}$
(or, $z_{f}$ and $\tau$).
 Fitting parameters are given in \citet{No07}.
 Data points correspond to measurements from \citet{Zh07}(open circles), 
 \citet{Ma07}(closed circles), \citet{Ig07}(open squares),
 \citet{Bu08}(closed squares), 
 \citet{Be07}(open upward triangles), \citet{Pa06}(closed upward
 triangles), \citet{Da09}
 (open diamonds), \citet{Ch09}(closed diamonds) and \citet{PG08}(open
 downward triangles).
} 
  \label{fig:3.1}
\end{figure}

The mass increase $\dot{M}_{all}$ is determined by
the star formation rate.
We adopt the staged $\tau $ model introduced by \citet{No07}
and use the same values and uncertainty ranges.
The main model parameters are $c_{\alpha}$ and 
$c_{\beta}$, which control the gas exhaustion timescale and the formation 
redshift, respectively.  
The model by \citet{No07} is consistent with observations over 
a wide range of masses and may be used as a well-fitting formulation 
of specific star formation rate (SSFR). Figure \ref{fig:3.1} shows 
the SSFR $\Psi /M$ at $z=$0.3, 0.5, 0.7 and 0.9. Here we include much 
more observational data, and obtain the same results as \citet{No07}.
For the stellar mass function data, we use the data 
of \citet{Il09} from the COSMOS
survey covering 2-deg$^{2}$, in which morphological and spectral
classifications were carried out. The combined classifications allow 
us to isolate the "blue elliptical" galaxies that are 
not included in the red sequence. We use the early-type (elliptical in 
\citet{Il09})
galaxy mass function at $z\le1.2$ as the spheroid mass function. 

\begin{figure}
 \begin{center}
  \includegraphics[width=110mm, angle=270]{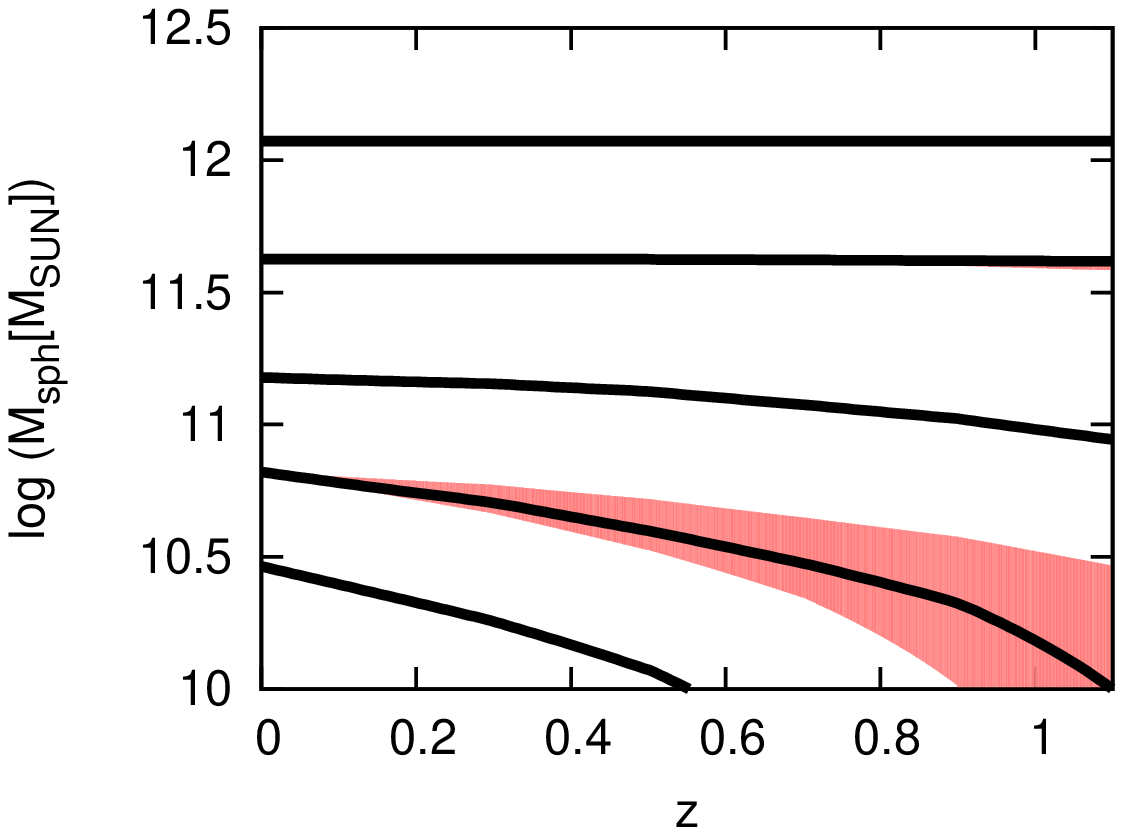}
 \end{center}
 \caption{Averaged growth history of spheroids computed using B/T=0.7 
  and SSFR with $ c_{\alpha}=10^{20.7}$ and $ c_{\beta}=10^{-2.7}$.
  Red shaded areas show the difference in $\log M_{sph}$ between 
  B/T=0.4 and B/T=1.0 for representative tracks at $z=$0.3, 0.5, 0.7, 0.9 and 1.1.}
 \label{fig:3.4}
\end{figure}

\begin{figure}
 \begin{center}
  \includegraphics[width=110mm, angle=270]{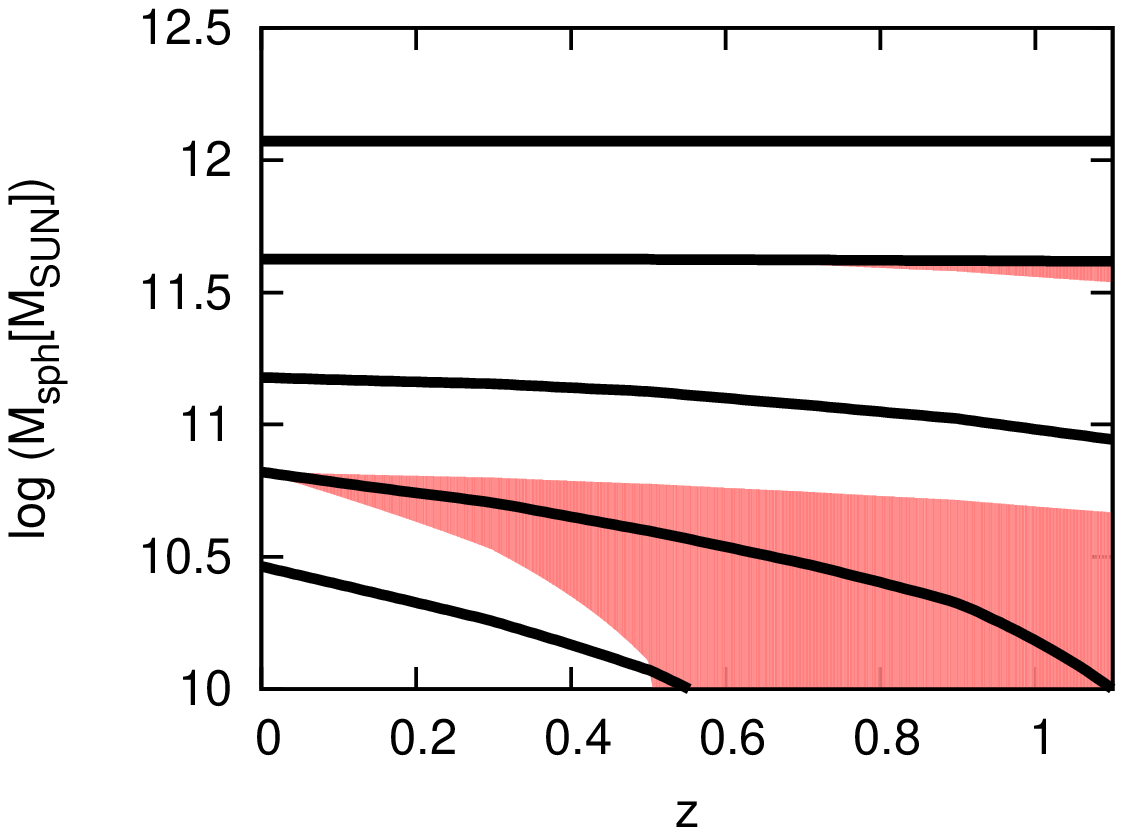}
 \end{center}
 \caption{Averaged growth history of spheroids computed using B/T=0.7 
  and SSFR parameters ($c_{\alpha}$, $c_{\beta}$) 
  with ($10^{20.7}$, $10^{-2.7}$).
   Red shaded areas show the difference in $\log M_{sph}$ between 
  ($c_{\alpha}$, $c_{\beta}$)=($10^{20.4}$, $10^{-1.7}$) and 
  ($10^{20.7}$, $10^{-3.0}$) for representative tracks at 
  $z=$0.3, 0.5, 0.7, 0.9 and 1.1.}
 \label{fig:3.5}
\end{figure}

The mass evolution is determined by easily integrating eq. (\ref{3.8})
since eq. (\ref{3.7}) is expressed as a smooth curve.
The results are shown in Figure \ref{fig:3.4}. 
It is found that there is no significant growth in 
the mass range of $  M \ge 10^{11} M_{\odot }$.
As discussed regarding the evolution of SMBHs,
uncertainty in the model parameters is
important in the phase of rapid growth. The parameters of
spheroid evolution are the B/T ratio, and the SSFR
parameters ($c_{\alpha}$ and $c_{\beta}$).
We use B/T=0.7, but examine the variation for 0.4$<$ B/T $<$1,
which \cite{Im02} considered as a reasonable range.
The range of variation is shown for
two evolutionary tracks in Figure \ref{fig:3.4}.
The uncertainty in $\log M_{sph} $ for
$M_{sph} = 10^{10.3}M_{\odot }$ at $z=0.9$
is $\sim 0.6$dex.
This uncertainty can be neglected for $M_{sph} > 10^{11}M_{\odot }$.
We also examine the variation due to the SSFR
parameters in Figure \ref{fig:3.5}.
It is found that the variation in the SSFR parameters affects the evolution more strongly
than that in the B/T ratio.
However, neither affects the spheroid
evolution for $M_{sph} > 10^{11.5}M_{\odot }$.

\section{ RESULTS }
    \subsection{Evolution of the $M_{BH}$-$M_{sph}$ relation}

\begin{figure}
 \begin{center}
  \includegraphics[width=110mm, angle=270]{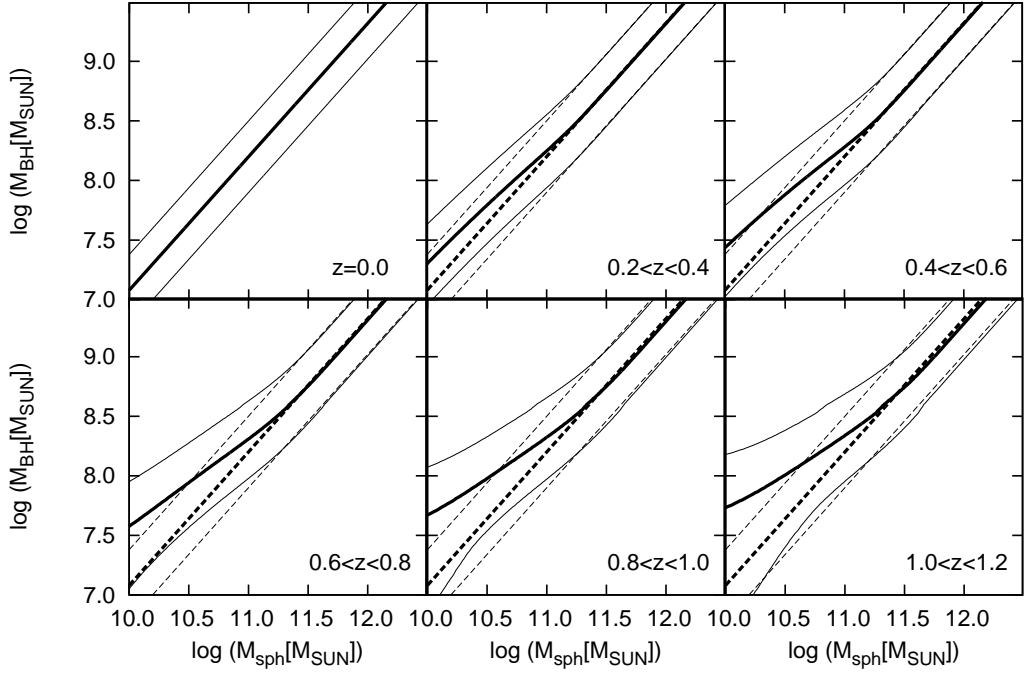}
 \end{center}
 \caption{Evolution of $M_{BH}$-$M_{sph}$ 
  relation from $z\sim1.2$ to $z=0$. 
  Model parameters are $\lambda=0.42$, B/T=0.7, 
  $ c_{\alpha}=10^{20.7}$ and $ c_{\beta}=10^{-2.7}$.
  An offset $\pm0.3$dex is added due to
 scatter in the local $M_{BH}$-$M_{sph}$ relation. 
 Relation, and upper and lower limits
 at $z=0$ are plotted by thick dotted lines and thin dotted lines, 
respectively.
}

 \label{fig:4.1}
\end{figure}

  The combination of models for SMBH mass and spheroid mass 
allows us to investigate the mass-dependent evolution of 
the $M_{BH}$-$M_{sph}$ relation.
In the local universe, 
the $M_{BH}$-$M_{sph}$ relation found by \citet{HR04} satisfies 
\begin{equation}
\log (M_{BH}/M_{\odot })=8.20+1.12\log (M_{sph}/10^{11}M_{\odot }).
\label{4.1}
\end{equation}
Using this, we consider the relation at $ 0 < z < 1.2$.
Figure \ref{fig:4.1} shows our results for a reference
model with $\lambda =0.42$, B/T=0.7, $c_{\alpha }=10^{20.7}$ and 
$c_{\beta}=10^{-2.7}$. 
In order to account for the intrinsic scatter at $z=0$, 
we add an offset of $\pm 0.3$dex \citep{HR04} in eq. (\ref{4.1}),
so that two curves corresponding to upper and lower limits
are also plotted at each value of $ z $.
We found that the $M_{BH}$-$M_{sph}$ relation does not change 
since $z\sim 1.1$ in the region for 
$M_{BH}$ $\simgt$ $10^{8.5}M_{\odot }$ 
and $M_{sph}$ $\simgt$ $10^{11.3}M_{\odot }$. 
The reason is clear in Figures \ref{fig:2.1} and \ref{fig:3.4},
where no evolution was seen in either mass on the massive side. 
On the other hand, 
there is a significant deviation 
from the linear relation when $M_{BH}<10^{8}M_{\odot }$. 
This denotes that 
a massive BH with $M_{BH}=10^{8}M_{\odot }$ is already 
located at a relatively small spheroid mass of
$M_{sph}=10^{10.5}M_{\odot }$ at $z \sim 1$.
The spheroid mass subsequently increases to  
$M_{sph}=10^{10.8}M_{\odot }$, while $M_{BH}$ is
fixed at $10^{8}M_{\odot }$.
In order to evaluate the increase of the SMBH mass
for a fixed spheroid mass $M_{sph}$, we consider
the difference 
$\Delta \log (M_{BH}) = \log (M_{BH}(z)/M_{BH}(0)) $.
When the evolution is plotted as $ \log (1+z)^{\gamma }$,
we obtain 
$\gamma =2.11$ at $M_{sph}=10^{10}M_{\odot }$,   
$\gamma=1.05$ at $M_{sph}=10^{10.5}M_{\odot}$ and
$\gamma =0.43$ at $M_{sph}=10^{11}M_{\odot }$. 
In the higher-mass regime of $M_{sph}>10^{11}M_{\odot }$, 
no significant evolution is found. 
Thus, the evolution of the $M_{BH}$-$M_{sph}$ relation
at $z $ $\simlt$ $1.1$ 
clearly depends on the mass range.
The growth is more rapid for a smaller spheroid mass.  
If we integrate the relation with the mass function over the 
whole mass range, then the averaged power index becomes $\gamma \sim0.8$.
We show the relation with $\gamma =0.8$ for comparison in subsection 4.2
 (Figure \ref{fig:4.4}). 
This parameterization for evaluating cosmological evolution 
has been used by \citet{MRD04}, and our result is
consistent with their value within an error of $1\sigma$.  
The index of the average is almost the same as that of the previous 
phenomenological model (e.g., \citet{MRD04}), but
mass-dependent evolution, especially 
at lower mass, is a remarkable new property in our model.

\begin{figure}
 \begin{center}
  \includegraphics[width=110mm, angle=270]{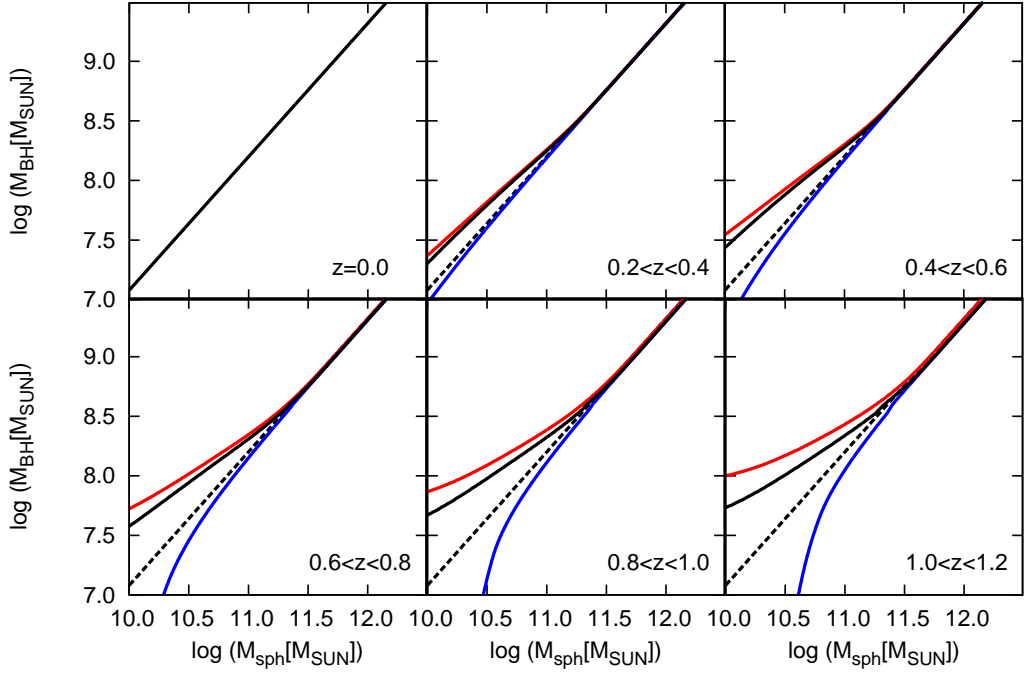}
 \end{center}
 \caption{
 Comparison of evolutionary models of $M_{BH}$-$M_{sph}$ relation.
 Black solid line is our reference model;
 black dotted line is relation at $z=0$.
 Model with fixed SMBH mass is denoted by top (red) line; 
 model with fixed spheroid mass is denoted by bottom (blue) line.}
 \label{fig:4.2}
\end{figure}

   In order to compare the growth of SMBHs and spheroids, 
we present two extreme cases in Figure \ref{fig:4.2}. 
One is calculated without evolution of SMBHs, and the other is
calculated without evolution of spheroids.
Both two curves deviate from our reference model as well as 
from the $M_{BH}$-$M_{sph}$ relation at $z=0$, as $z$ increases.
The deviation is remarkable at lower mass.
However, our reference model is close to the curve for 
no evolution of SMBHs.
This means that the mass increase of the spheroids around 
$M_{sph}  = 10^{10.5} M_{\odot}$ is more rapid.

\begin{figure}
 \begin{center}
  \includegraphics[width=110mm, angle=270]{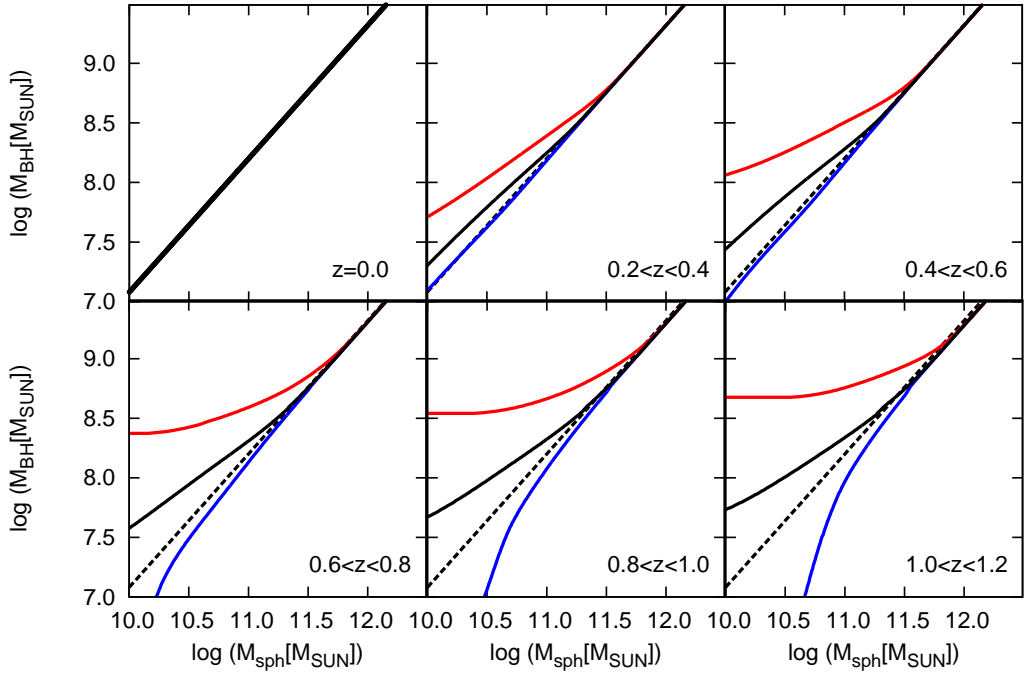}
 \end{center}
 \caption{ Evolution of $M_{BH}$-$M_{sph}$ relation from
 $z\sim1.2$ to $z=0$ varying model parameters. 
 Black solid line is our reference model;
 black dotted line is $M_{BH}$-$M_{sph}$ relation at $z=0$.
 Top (red line) is upper limit, for which $\lambda=0.28$, 
 B/T=0.4, $ c_{\alpha}=10^{20.4}$ and $ c_{\beta}=10^{-1.7}$;
 Bottom (blue line) is lower limit, for which 
 $\lambda=0.56$, B/T=1.0, $ c_{\alpha}=10^{20.7}$ 
 and $ c_{\beta}=10^{-3.0}$.  }
 \label{fig:4.3}
\end{figure}

We examine the model parameters discussed in Section 2 and 3 to 
demonstrate deviation from the reference model.
The three parameters are independent and the results change
monotonically as each parameter is changed.
Thus, we have calculated two extreme cases: upper and lower limits
of the $M_{BH}$-$M_{sph}$ relation.
These results are shown in Figure \ref{fig:4.3}.
For the model with $c_{\alpha} = 10^{20.7} $ and 
$c_{\beta}=10^{-3.0}$, which corresponds to the curve for the upper
limit in Figure \ref{fig:4.3}, 
the distribution of $M_{BH} $ 
in the range of $M_{sph} $ $\simlt$ $10^{11}M_{\odot}$ 
is a nearly horizontal line at $z\sim1.1$.
On the other hand,
for $c_{\alpha}=10^{20.4}$ and $c_{\beta}=10^{-1.7}$,
corresponding to the curve for the lower limit in Figure \ref{fig:4.3},
the distribution in the same range is even lower than 
that of the local relation. 
For lower masses, the deviation from the linear line
is general, and the typical SMBH mass depends on the model.
For example, our model shows that $M_{BH}$ ranges 
from $10^{7}M_{\odot}$ to $10^{8.5}M_{\odot}$
for $M_{sph} = 10^{10.5}M_{\odot}$ at $z=1$.
The possible range of masses is wide, but
our reference model, which can be regarded as an average of 
two extreme cases, suggests slightly larger SMBH mass than 
that of the local universe relation. 
 It is also found that the relation is not affected 
in the mass range of $M_{sph}$ $\simgt$ $10^{11.5}M_{\odot}$.

\subsection{Comparison with observations}

\begin{figure}
 \begin{center}
  \includegraphics[width=110mm, angle=270]{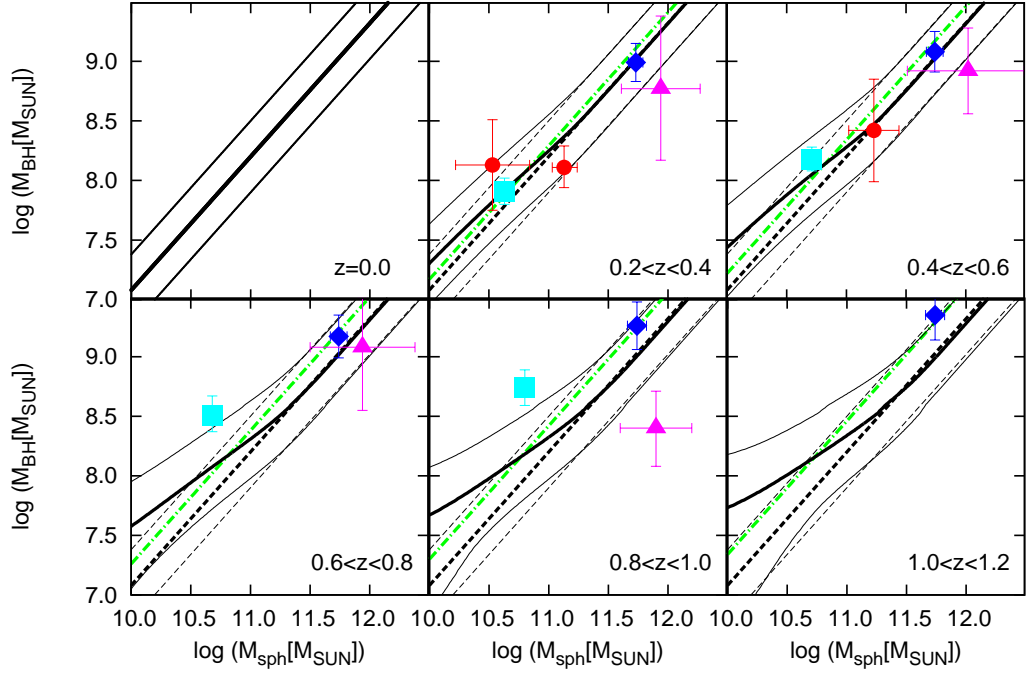}
 \end{center}
 \caption{ Comparison of $M_{BH}$-$M_{sph}$ relation from
 $z\sim1.2$ to $z=0$ with observations. 
 Solid lines are reference models and dotted 
 lines are local relations.
 The relation shifted by $(1+z)^{0.8}$ is shown by dash-dotted (green)
 lines.
 Red circles: \citet{Be09}; light 
 blue squares: 
 \citet{Sa07}; purple triangles: \citet{De09}; blue 
 diamonds: \citet{Mc06}. 
 Error bars are standard deviation of the logarithmic mean.}
 \label{fig:4.4}
\end{figure}

 In this subsection, we compare our model with observational 
results available in the literature.
\citet{Mc06} investigated the evolution of the SMBH-to-host galaxy mass 
ratio in the redshift range of $0<z<2$.
They estimated the masses $M_{BH}$ for the 3C RR quasars and 
spheroid masses $M_{sph}$ for the 3C RR radio galaxies. 
Two samples are not the same, but the averaged mass was 
determined. 
Recently, \citet{De09} studied the redshift dependence of 
the relation up to $z=3$ using a sample of 96 quasars 
for which the host galaxy luminosity is known. 
They listed $M_{BH}$ and $M_{sph}$ for each source.
Both results are used for comparison after we derive
$M_{BH}$ from the observational data by the same method. 
Black hole masses are usually estimated by the virial 
method (e.g., \citet{MJ02, Ko06}).
The masses are derived from the broad emission-line width $v$ 
as a velocity indicator, 
and the monochromatic continuum luminosity $L$ is used as an indicator of 
the region size:
\begin{equation}
 \log M_{BH} = A + 2\log v + B\log L,
\end{equation}
where $A$ and $B$ are constants.
The line width $v$ is typically taken from the broad line
of Mg$_{II}$ or H$\beta$, and the luminosity $L$
is taken at 3000\AA  or 5100\AA.
The constants $A$ and $B$ have not been determined with high accuracy, since the best fit values
slightly depend on observational choices of $v$ and $L$.
The black hole masses are derived by a different choice of
$A$ and $B$ in the works of \citet{Mc06} and \citet{De09}.
It is preferable to use a single formula for the estimation.
One possible approach is to use the values, $A$ and $B$, 
derived by \citet{Mc08}, who cross-calibrated a number of different 
formulae in a redshift range where more than one broad emission 
line could be observed simultaneously in optical spectra.
We adopt their calibrated values for $A$ and $B$,
in order to re-derive black hole masses. 
We calculate $M_{BH}$ of the objects in 
the samples by \citet{Mc06} and \citet{De09},
for which the line width of Mg$_{II}$ or H$\beta$ is measured.
The corresponding line width of nine objects is not listed; consequently,
these are excluded from our analysis.
In Figure \ref{fig:4.4}, we plot $(M_{BH}, M_{sph})$ not 
for each source but for the average,  
since we focus only on statistical properties. 
Our values $M_{BH}$ and those listed in 
\citet{Mc06} and \citet{De09} are not notably different by averaging.
\citet{Be09} investigated the evolution of the $M_{BH}$-$L_{sph}$
relation.
\footnote{
Similar samples were previously given by \cite{Tr07}, but
most of these were covered by the sample of \cite{Be09}.} 
As discussed previously, $M_{BH}$ is derived 
from spectral data 
using the mass estimator by \citet{Mc08}.
The spheroid mass $M_{sph}$ is derived  
from $L_{sph}$ and the mass-to-light ratio 
by a certain stellar population model(\cite{BC03}). 
Thus, converted data are also plotted in Figure \ref{fig:4.4}. 
We group data into two ranges, 
$M_{sph}<10^{11}M_{\odot}$ and $M_{sph}\ge 10^{11}M_{\odot}$,
and respectively average $M_{BH}$ and $M_{sph}$, since
the sample number is small.
Although detailed comparison is difficult
due to the small sample and the ambiguity in the conversion relation, 
the results are consistent with mass-dependent evolution. 
The samples are deawn from the high mass end of the spheroid mass function 
($M_{sph}>10^{11.5}M_{\odot}$), where we expect little or no evolution in the 
$M_{BH}$-$M_{sph}$ relation since $z\sim1$. 
Comparing our evolutionary model with observations in each redshift bin we 
find consistency with observations, although we cannot exclude a mass-independent 
scenario because of the lack of observational data in the low spheroid mass range. 

  We next consider additional observation data, although the 
comparison becomes indirect.
  \citet{Sa07} investigated the evolution of the 
$M_{BH}$-$\sigma$ relation.
An evolutionary model of $\sigma$ is necessary, but
here we assume that the conversion relation 
from $\sigma$ to $M_{sph}$ is the same as that of $z=0$.
In other words, the relation between 
$\sigma$ and $M_{sph}$ at $z=0$ is derived by eliminating $M_{BH}$
in eq. (\ref{4.1}) and the $M_{BH}$-$\sigma$ relation at $z=0$ 
by \citet{Tr02}, and the relation is assumed to hold for all $z$.  
Averaged data points from \citet{Sa07} are also plotted 
in Figure \ref{fig:4.4}. 
These data correspond to small masses of 
$M_{sph} <10^{11} M_{\odot} $ and 
show relatively strong evolution, 
supporting the hypothesis of a mass-dependent evolution scenario.

\section{SUMMARY AND CONCLUSIONS}

  We have modeled the mass-dependent evolution of black holes and 
spheroids since $z\sim 1.2$. The black hole mass evolution was investigated
by using a continuity equation and the observed AGN luminosity function. 
The spheroid mass evolution was also derived from a continuity equation and 
the observed SSFR. Both evolutions, which were 
consistent with "downsizing", significantly depend on 
the masses of the SMBHs and spheroids, and the $M_{BH}$-$M_{sph}$ relation 
is mass-dependent.
We assumed that all galaxies at $z=0$ satisfy
the linear relation of the $\log(M_{BH})$-$\log(M_{sph})$ diagram
with some uncertainty, and examined the history to $z\sim1.2$. 
It was found that the relation is unchanged 
for $0< z <1.2$ in the range of
$M_{BH}>10^{8}M_{\odot}$ and $M_{sph}>10^{11}M_{\odot}$.
In the low-mass regime, however, there is clear deviation 
from the linear relation.
The mass increase of the spheroids is likely larger than 
that of black holes, although there is uncertainty in
the model parameters.
Our model suggests that SMBHs were already located in spheroids with 
relatively small mass, compared with the local relation.  

We also compared our results with currently available 
observations concerning the relations between $M_{BH}$ and 
host spheroid parameters at certain cosmic ages.
Our mass-dependent evolutionary model is consistent with 
these observations, although the observational sample 
is not sufficiently complete to make a statistically significant 
comparison. 
Larger observational samples are needed to study the history of 
SMBHs and their hosts statistically, expecially in the low-mass regime 
where the differences between the mass-dependent and the mass-independent 
evolution models are higher. 
If a break from the linear $\log(M_{BH})$-$\log(M_{sph})$ is found, our 
mass-dependent scenario would be supported and it would be possible to put 
stronger constraints on model parameters. 
It would also be of great interest to extend the present results to higher 
redshift, where the model uncertainties are considerably smaller. 
This would enable us to reveal possible inconsistencies between model and observations. 
However, the extension of the present study to higher redshift is not straightforward, 
as the adopted stellar mass function for early-type galaxies is only significant 
at $z<1.2$ \citep{Il09}.

\clearpage

%\input{fig}
%Fig.1%
\clearpage
%Fig.2%
\begin{figure}
 \begin{minipage}{0.47\hsize}
  \begin{center}
   \includegraphics[width=47mm, angle=270]{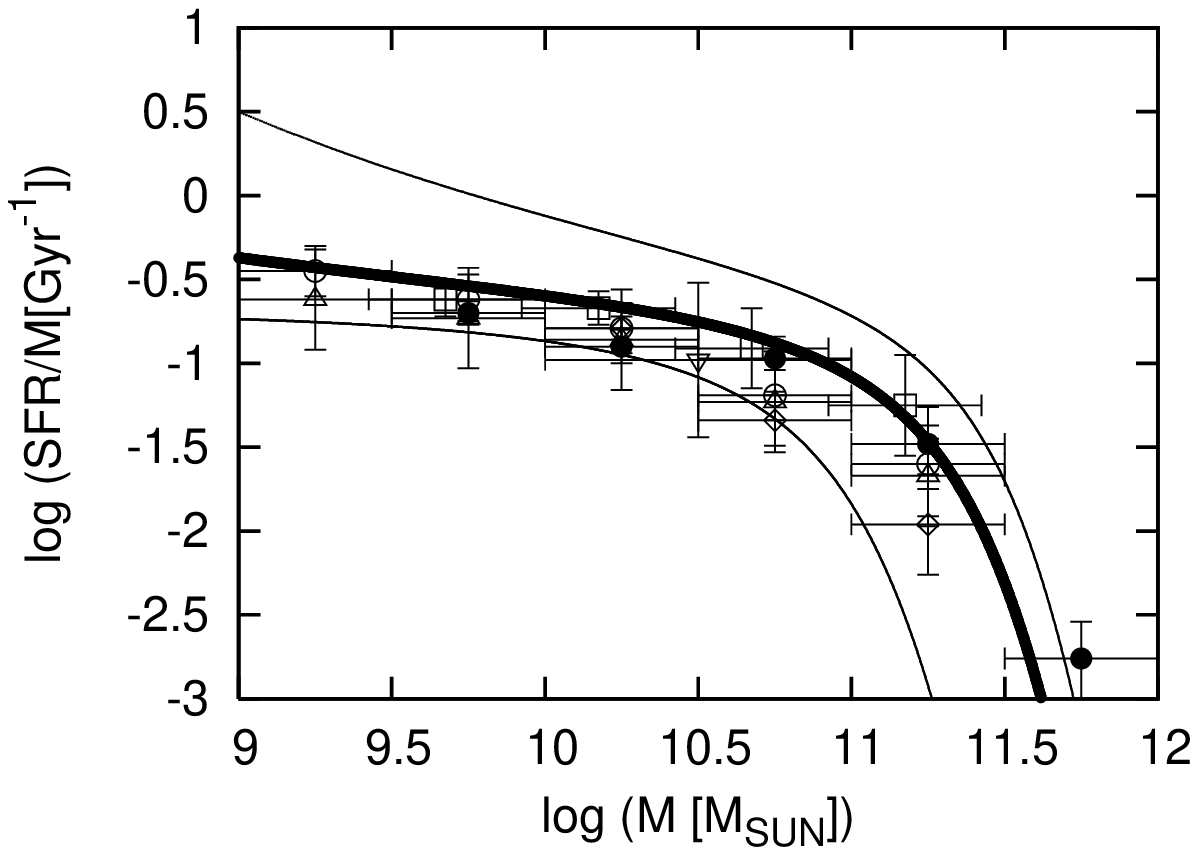}
  \end{center}
 \end{minipage}
 \begin{minipage}{0.47\hsize}
  \begin{center}
   \includegraphics[width=47mm, angle=270]{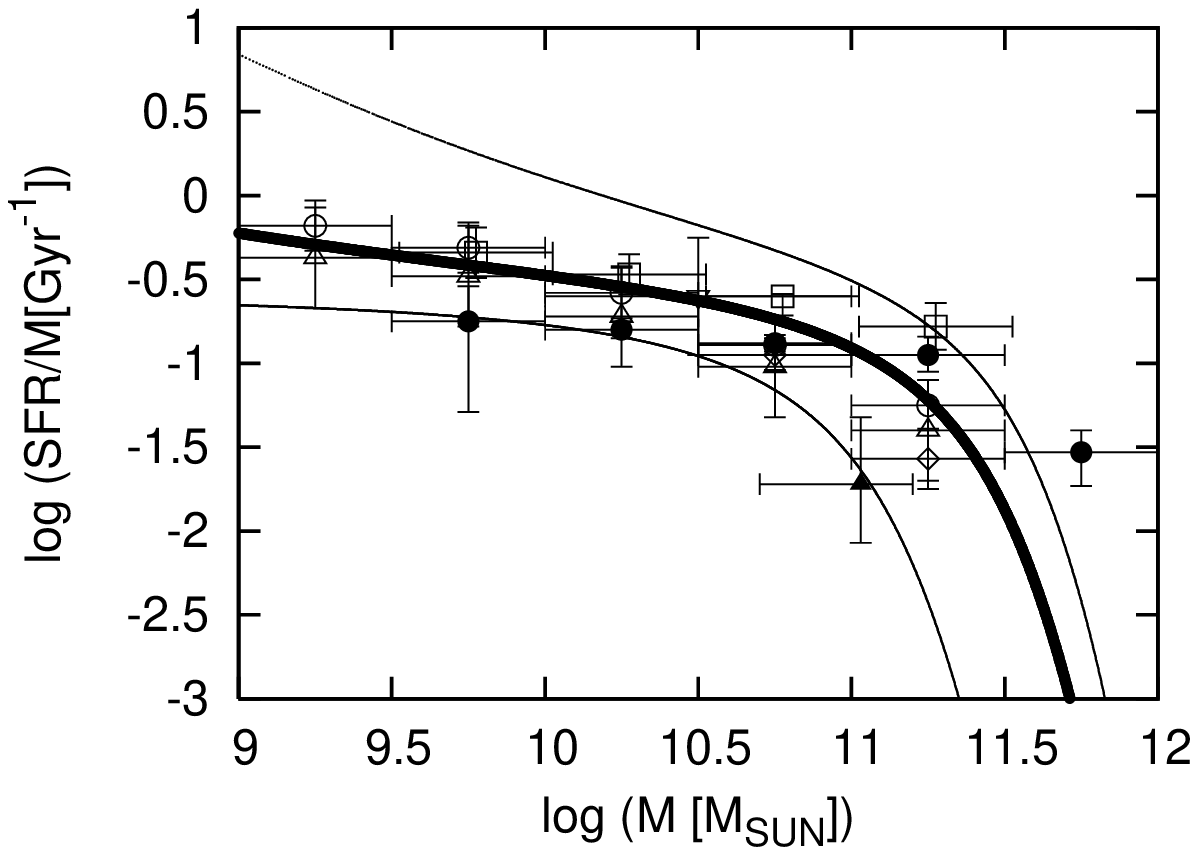}
  \end{center}
 \end{minipage}
\end{figure}
%Fig.3%

\clearpage

%Fig.4%

%Fig.5%

\clearpage

%Fig.6%

\clearpage

%Fig.7%

\clearpage
%Fig.8%

\end{document}